\documentclass[aps,pra,preprint,superscriptaddress,showpacs]{revtex4-1}

\usepackage{graphicx}
\usepackage{color}

\newcommand {\bfr} {{\bf r}}

\newcommand {\om} {\omega}

\begin{document}

\title{Quantum and classical features of the photoionization spectrum of C$_{60}$}

\author{A.V. Verkhovtsev}
\email[]{verkhovtsev@fias.uni-frankfurt.de}
\affiliation{Frankfurt Institute for Advanced Studies, Ruth-Moufang-Str. 1, 60438 Frankfurt am Main, Germany}
\affiliation{St. Petersburg State Polytechnic University, Politekhnicheskaya ul. 29, 195251 St. Petersburg, Russia}
\affiliation{A.F. Ioffe Physical-Technical Institute, Politekhnicheskaya ul. 26, 194021 St. Petersburg, Russia}

\author{A.V. Korol}
\affiliation{Frankfurt Institute for Advanced Studies, Ruth-Moufang-Str. 1, 60438 Frankfurt am Main, Germany}
\affiliation{Department of Physics, St. Petersburg State Maritime Technical University,
Leninskii prospekt 101, 198262 St. Petersburg, Russia}

\author{A.V. Solov'yov}
\altaffiliation{On leave from A.F. Ioffe Physical-Technical Institute, Politekhnicheskaya ul. 26,
194021 St. Petersburg, Russia}
\affiliation{Frankfurt Institute for Advanced Studies, Ruth-Moufang-Str. 1, 60438 Frankfurt am Main, Germany}

\date{\today}

\begin{abstract}
By considering photoionization of the C$_{60}$ fullerene, we elucidate the contributions of various
classical and quantum physics phenomena appearing in this process.
By comparing the results, based on the {\it ab initio} and model approaches, we map the well-resolved
features of the photoabsorption spectrum to single-particle and collective excitations which have the
different physical nature.
It is demonstrated that the peculiarities arising in the photoionization spectrum of C$_{60}$ atop the
dominating plasmon excitations have the quantum origin.
In particular, we demonstrate that a series of individual peaks can be assigned either to the optically
allowed discrete transitions or to ionization of particular molecular orbitals of the system.
The analysis performed gives a detailed information on the nature of numerous features in the
photoabsorption spectrum of C$_{60}$.
\end{abstract}



\maketitle


\section{Introduction}

Dynamics of electron excitations and dynamical processes of fullerenes and other carbon-based
nanoscale systems have been the topic of intensive experimental and theoretical research for more
than several decades (see, e.g., \cite{Sattler_Handbook_of_Nanophysics_2,
Sattler_Handbook_of_Nanophysics_4, Solovyov_review_2005_IntJModPhys.19.4143} for review).
A special attention has been paid to the study of ionization processes of the systems by means of
the photon, electron and ion impact
\cite{Solovyov_review_2005_IntJModPhys.19.4143, Bertsch_1991_PhysRevLett.67.2690,
Hertel_1992_PhysRevLett.68.784, Berkowitz_1999_JChemPhys.111.1446, Reinkoester_2004_JPhysB.37.2135, Scully_2005_PhysRevLett.94.065503, Kafle_2008_JPhysSocJpn.77.014302,
Keller_Coplan_1992_ChemPhysLett.193.89, Mikoushkin_1998_PhysRevLett.81.2707,
Verkhovtsev_2012_JPhysB.45.141002}.

Photoionization of fullerenes, as well as other nanoscale systems, represents a complex phenomenon
and involves a number of features which can be studied by means of various theoretical methods.
Being by its nature a quantum phenomenon, the photoionization process can be described within the
{\it ab initio} framework based on the time-dependent density-functional theory (TDDFT)
\cite{Runge_Gross_1984_PhysRevLett.52.997}.
It deals with the time-dependent Schr\"{o}dinger equation and allows one to obtain information on the
excited-state properties of a complex many-electron system.
However, it is well established that photoionization of nanoscale carbon systems, fullerenes in particular,
as well as various metallic clusters and nanoparticles, takes place through plasmons --- collective
excitations of delocalized valence electrons which are induced by an external electric field.
The plasmon excitations correspond to oscillations of the electron density with respect to the
positively charged ions \cite{Lundqvist_Electron_Gas, Connerade_AS_PhysRevA.66.013207}.
Such collective excitations, appearing in many-electron systems,
are well known in classical electrodynamics and are described in the classical physics terms
\cite{Lundqvist_Electron_Gas, Connerade_AS_PhysRevA.66.013207, Verkhovtsev_2012_EPJD.66.253}.

When a fullerene is ionized either by a photon or by a projectile particle, various types of collective
excitations, which are characterized by prominent resonant-like structures in the ionization spectra,
are formed in the system.
The most prominent structure, positioned in the excitation energy range from 20 to 30 eV, is formed
due to collective oscillations of both $\sigma$ and $\pi$ delocalized electrons of a system, while
a smaller narrow peak in the low-energy region of the spectrum (below 10 eV) is attributed to the
collective excitation of only $\pi$-electrons.
The $\sigma$- and $\pi$-electrons occupy, respectively, $\sigma$- and $\pi$-orbitals of a fullerene,
which are formed due to the sp$^2$-hybridization of carbon atomic orbitals.
The resonance peaks in the ionization spectra are described by some characteristic widths, $\Gamma$,
which have a quantum origin and appear due to the decay of the collective excitation modes into the
incoherent sum of single-electron excitations.
Although the exact calculation of the plasmon widths should be performed within the quantum-mechanical
framework, they can be estimated using the relation similar to the Landau damping of plasmon oscillations
\cite{Solovyov_review_2005_IntJModPhys.19.4143}.
Such an estimate results in $\Gamma \sim A v_F /R$, where $v_F$ is the velocity of the fullerene
electrons on the Fermi surface, $R$ describes a characteristic width of the electron density distribution
in the fullerene, and $A$ is a factor on the order of unity.

In most cases, the excitation spectra, calculated within the {\it ab initio} framework, can be obtained
in a broad range of excitation energies range only for small molecules or clusters consisting of a few atoms.
For larger system, such as, for instance, fullerenes, a vast majority of contemporary software packages
for {\it ab initio} based calculations can describe accurately only a limited number of low-lying excited
states located below or just above the ionization threshold.
A detailed structure of the spectrum at higher excitation energies, where the plasmon excitations dominate
the spectrum, could be hardly revealed due to significant computational costs.
An alternative approach for the description of electron excitations in many-electron systems is based
on the jellium model \cite{Ekardt_1984_PhysRevB.29.1558}.
During the recent years, this approach has been used to calculate the photoionization
spectra of fullerenes and atoms, endohedrally confined inside the fullerene cages
(see, e.g. \cite{Solovyov_review_2005_IntJModPhys.19.4143, Belyaev_2007_IntJQuantChem.107.2781,
Madjet_2008_JPhysB.41.105101, Maser_2012_PhysRevA.86.053201}).

An effective tool for evaluation of the contribution of plasmon excitations to the ionization spectra
is based on the plasmon resonance approximation
\cite{Gerchikov_1997_JPhysB.30.4133, Gerchikov_1998_JPhysB.31.3065, Gerchikov_1997_JPhysB.30.5939}.
The advantage of this approach is that it provides a clear physical explanation of the resonant-like
structures in the photoionization \cite{Connerade_AS_PhysRevA.66.013207, Korol_AS_2007_PhysRevLett_Comment}
and inelastic scattering cross sections
\cite{Gerchikov_1997_JPhysB.30.4133, Gerchikov_1998_JPhysB.31.3065, Gerchikov_1997_JPhysB.30.5939,
Verkhovtsev_2012_JPhysB.45.141002, Bolognesi_2012_EurPhysJD.66.254}
on the basis of excitation of plasmons by the photon or electron impact.

In this paper, we elucidate the contributions of various classical and quantum physics phenomena appearing
in the ionization process.
By comparing the {\it ab initio} TDDFT results with those based on the plasmon resonance approximation,
we map the well-resolved features of the photoabsorption
spectrum of the C$_{60}$ fullerene to different types
of single-particle and collective electron excitations having the different physical nature.
We demonstrate that the peculiarities arising in the spectrum atop the dominating plasmon excitations
have the quantum origin.
In particular, we demonstrate that a series of individual peaks in the continuous part of the excitation
spectrum can be assigned to the particular single-electron transitions and
caused by ionization of inner molecular orbitals of the fullerene.
%
It is also demonstrated that the results of the {\it ab initio} and model-based calculations are in close
agreement with experimental results on photoabsorption of C$_{60}$.
To our knowledge, this is the first study which presents a detailed analysis of the photoionization
spectrum of such a widely explored system as C$_{60}$.

The atomic system of units, $m_e = |e| = \hbar = 1$, is used throughout the paper.

\section{Theory}

\subsection{Time-dependent density-functional theory}

In the present study we utilize the TDDFT approach to calculate precisely the photoabsorption
spectrum of C$_{60}$.
Being a generalization of density-functional theory (DFT) \cite{Kohn_Sham_1965_PhyRev.140.A1133},
TDDFT allows one to introduce time-dependent Kohn-Sham equations \cite{Runge_Gross_1984_PhysRevLett.52.997}
and to study various single-particle properties of a many-electron system as a function of time.
Within the TDDFT approach the response function of the system can be calculated either in the time or in
the frequency domain.
In the former case, one uses the real-time propagation method
\cite{Yabana_Bertsch_1996_PhysRevB.54.4484, Kawashita_2009_JMolStruct_Theochem.914.130}
to study the evolution of the dipole moment due to an initial impulsive distortion of the system.
The main limitation of this approach is that stable integration of the time-dependent Kohn-Sham
equations requires a very small time step, $\sim 10^{-3}$ fs, which decreases with increasing the number
of grid points \cite{Walker_2006_PhysRevLett.96.113001} and, therefore, is very demanding from a
computational viewpoint.
Within the alternative method \cite{Walker_2006_PhysRevLett.96.113001, Rocca_2008_JChemPhys.128.154105}
based on the frequency representation of the response function, it is possible to calculate the full
photoabsorption
spectrum in a broad energy range without repeating time-consuming operations for different excitation
frequencies.
In this approach, the response function is represented by a matrix element of the resolvent of the
Liouvillian operator (see \cite{Walker_2007_JChemPhys.127.164106, Rocca_2008_JChemPhys.128.154105} for details).
This approach has been used recently to study photoionization of noble gas atoms encapsulated
in C$_{60}$ \cite{Chen_Msezane_2012_EPJD.66.184, Chen_Msezane_2012_PhysRevA.86.063405}.

In the present study the {\it ab initio} TDDFT calculations are performed in the linear regime within the
dipole approximation.
The linear response theory aims to study the variation of a given physical observable
due to the application of a weak external perturbation to a many-particle system.
Within this framework, the external potential acting on the system can be represented as a sum of a
time-independent part, $v_{\rm ext}^0(\bfr)$, and a time-dependent perturbation,
$v_{\rm ext}^{\prime}(\bfr, t)$:
\begin{equation}
v_{\rm ext}(\bfr,t) = v_{\rm ext}^0(\bfr) + v_{\rm ext}^{\prime}(\bfr,t) \ .
\label{TDDFT_eq_04}
\end{equation}

\noindent Application of the external perturbation leads to variation of the electron density of the system.
Therefore, the time evolution of the electron density can be represented as a sum of two terms,
$\rho(\bfr, t) = \rho_0(\bfr) + \delta\rho(\bfr,t)$,
where $\rho_0(\bfr)$ is the unperturbed ground-state density, and $\delta\rho(\bfr, t)$
describes variation of the electron density due to the perturbation $v^{\prime}_{\rm ext}(\bfr, t)$.

In order to consider the response of a system to an external perturbation in the frequency representation,
one performs the Fourier transformation of time-dependent quantities.
In the linear regime, the Fourier transform of $\delta\rho(\bfr, t)$ reads
\begin{equation}
\delta\rho(\bfr,\om) = \int \chi(\bfr,\bfr^{\prime},\om) v_{\rm ext}^{\prime}(\bfr^{\prime},\om)
{\rm d}\bfr^{\prime} \ ,
\label{TDDFT_eq_09}
\end{equation}

\noindent where $v_{\rm ext}^{\prime}(\bfr^{\prime},\om)$ is the Fourier transform of the external
perturbation $v^{\prime}_{\rm ext}(\bfr, t)$ and $\chi(\bfr,\bfr^{\prime},\om)$ is the generalized
frequency-dependent susceptibility of the system.

For the external perturbation $v_{\rm ext}^{\prime}(\bfr,\om) = - {\bf E}(\om) \cdot {\bfr}$
due to a uniform electric field, the Fourier transform of the induced dipole moment reads as follows:
\begin{equation}
d_i(\om) = \sum_j \alpha_{ij}(\om) E_j(\om) \ ,
\end{equation}

\noindent where $i, j$ denote the Cartesian components, $\alpha_{ij}(\om)$ is the dynamical polarizability
tensor which describes the linear response of the dipole to the external electric field:
\begin{equation}
\alpha_{ij}(\om) =
- \int r_i \chi(\bfr,\bfr^{\prime},\om) r_j^{\prime} \, {\rm d}\bfr {\rm d}\bfr^{\prime} \ ,
\end{equation}

\noindent and $r_i$ and $r_j^{\prime}$ are the components of the position operators $\bfr$ and
$\bfr^{\prime}$.
The photoabsorption cross section is related to the imaginary part of $\alpha_{ij}(\om)$ through
\begin{equation}
\sigma(\om) = \frac{4\pi\om}{3c} \sum_j {\rm Im} \alpha_{jj}(\om) \ ,
\end{equation}

\noindent where $c$ is the speed of light, and the summation is performed over the diagonal elements
of the polarizability tensor.

Within the approach introduced in \cite{Walker_2006_PhysRevLett.96.113001, Rocca_2008_JChemPhys.128.154105},
the electron density variation, $\delta\rho(\bfr,\om)$, is expressed via the so-called Liouvillian
operator $\mathcal{L}$,
\begin{equation}
(\om - \mathcal{L}) \cdot \delta\rho(\bfr,\om) =
\left[ v_{\rm ext}^{\prime}(\bfr,\om), \rho_0 \right] \ ,
\end{equation}

\noindent whose action onto $\delta\rho(\bfr,\om)$ is defined as:
\begin{equation}
\mathcal{L} \cdot \delta\rho(\bfr,\om) =
\left[ H_0, \delta\rho(\bfr,\om) \right] +
\left[ v^{\prime}_{\rm H}(\bfr,\om), \rho_0 \right] +
\left[ v^{\prime}_{\rm xc}(\bfr,\om), \rho_0 \right]\ ,
\end{equation}

\noindent where $H_0$ is the ground-state Kohn-Sham Hamiltonian calculated within the DFT approach,
and $v^{\prime}_{\rm H}(\bfr,\om)$ and $v^{\prime}_{\rm xc}(\bfr,\om)$ stand for the linear variations
of the frequency-dependent electrostatic and exchange-correlation potentials, respectively
\cite{Rocca_2008_JChemPhys.128.154105}.
The polarizability tensor $\alpha_{ij}(\om)$ is defined then by the off-diagonal matrix element
of the resolvent of the Liouvillian $\mathcal{L}$:
\begin{equation}
\alpha_{ij}(\om) = - \langle r_i | \left( \om - \mathcal{L} \right)^{-1} \cdot [r_j, \rho_0] \rangle \ ,
\end{equation}

\noindent which is calculated using the Lanczos recursion method
(see \cite{Rocca_2008_JChemPhys.128.154105,Walker_2007_JChemPhys.127.164106} for details).

As it was stressed above, the advantage of the utilized approach is that it makes possible to calculate
the full photoabsorption spectrum of complex molecular systems in a broad range of excitation energies.
However, this approach does not allow one to get information on partial ionization cross sections which
describe ionization of particular molecular orbitals.
The reason of this drawback is that only the occupied states are required for performing the calculations
and there is no need to calculate any empty states \cite{Walker_2006_PhysRevLett.96.113001, Rocca_2008_JChemPhys.128.154105}.
It makes the method, introduced in the aforementioned references, substantially different from
the Casida's approach \cite{Casida_RecentAdvances}, which is implemented in many codes for
{\it ab initio} calculations.
Within the latter one, it is possible to calculate each individual excitation and to assign it to a
specific transition.
In general, this operation is feasible only in a limited range of excitation energies, which is not
typically larger than 10 eV and depends also on the density of the excitation energies
\cite{Rocca_2007_Thesis}.
Alternatively, the method, introduced in Ref. \cite{Walker_2006_PhysRevLett.96.113001, Rocca_2008_JChemPhys.128.154105,Walker_2007_JChemPhys.127.164106}, allows one to compute the absorption
spectrum in a broad energy range but a systematic way to assign the transitions is missed.

\subsection{Plasmon resonance approximation}

In order to evaluate the contribution of plasmon excitations to the cross section, we utilize the
following model approach.
The fullerene is represented as a spherically symmetric system with a homogeneous charge distribution
within the shell of a finite width, $\Delta R = R_2 - R_1$, where $R_{1,2}$ are the inner and the
outer radii of the molecule, respectively
\cite{Verkhovtsev_2012_JPhysB.45.141002, Oestling_1993_EurophysLett.21.539,
Lambin_Lukas_1992_PhysRevB.46.1794, Lo_2007_JPhysB.40.3973}.
The chosen value of the shell's width, $\Delta R = 1.5$~\AA, corresponds to the typical size of the
carbon atom \cite{Oestling_1993_EurophysLett.21.539}.

Due to interaction with the uniform external field, $\bf{E}(\om)$, the variation of the electron density,
$\delta \rho(\bfr,\om)$, occurs on the inner and outer surfaces of the fullerene shell.
This variation leads to the formation of the surface plasmon, which has two normal modes,
the symmetric and antisymmetric ones \cite{Oestling_1993_EurophysLett.21.539,
Lambin_Lukas_1992_PhysRevB.46.1794, Lo_2007_JPhysB.40.3973, Korol_AS_2007_PhysRevLett_Comment}.
It was shown
\cite{Connerade_AS_PhysRevA.66.013207, Verkhovtsev_2012_EPJD.66.253, Korol_AS_2007_PhysRevLett_Comment},
that only the surface plasmon can occur in the system interacting with a uniform external electric field,
as it happens in the photoionization process.
When a system interacts with a non-uniform electric field created, for instance, in collision with
charged particles, the volume plasmon \cite{Gerchikov_2000_PhysRevA.62.043201} can also occur due to a
local compression of the electron density in the shell interior \cite{Bolognesi_2012_EurPhysJD.66.254}.

Within the plasmon resonance approximation (PRA)
\cite{Gerchikov_1997_JPhysB.30.4133, Gerchikov_1998_JPhysB.31.3065, Gerchikov_1997_JPhysB.30.5939}
it is assumed that the main contribution to the cross section
comes from the collective electron excitations.
Single-particle excitations are not accounted for in the approximation, since the single-particle
effects give a small contribution as compared to the collective modes
\cite{Gerchikov_1997_JPhysB.30.5939, Gerchikov_2000_PhysRevA.62.043201}.
Within this approach the dynamical polarizability $\alpha(\om)$ has a resonance behavior in the region
of frequencies where collective electron modes in a fullerene can be excited.
In the present study, we account for the both $\pi$- and $(\sigma+\pi)$-plasmons, which involve only
$\pi$ or both $\sigma+\pi$ delocalized electrons of the system, respectively.
Thus, the photoionization cross section, $\sigma_{\rm pl}(\om) \propto {\rm Im} \,\alpha(\om)$, is
defined as a sum of the two plasmons,
$\sigma_{\rm pl}(\om) = \sigma^{\pi}(\om) + \sigma^{\sigma+\pi}(\om)$,
and the contribution of each plasmon is governed by the symmetric and antisymmetric modes:
\begin{equation}
\sigma^i(\om) = \frac{4\pi \om^2}{c}
\left(
{ N_s^i \,\Gamma_s^i \over \bigl(\om^2-(\om_{s}^i)^2\bigr)^2+ \om^2(\Gamma_s^i)^2} +
{ N_a^i\,\Gamma_a^i \over \bigl(\om^2-(\om_{a}^i)^2\bigr)^2+ \om^2(\Gamma_a^i)^2}
\right) \ ,
\label{CS_plasmon}
\end{equation}

\noindent where the superscript $i$ denotes the $\pi$- or $(\sigma+\pi)$-plasmon.
Here $\om$ is the photon energy, $\om_s^i$ and $\om_a^i$ are, respectively, the resonance
frequencies of the symmetric and antisymmetric modes of the two plasmons,
$\Gamma_s^i$ and $\Gamma_a^i$ are the corresponding widths of the plasmon excitations,
$N_s^i$ and $N_a^i$ are the number of delocalized electrons which are involved in each collective excitation.
These values should obey the sum rule
$N_s^{\sigma+\pi} + N_a^{\sigma+\pi} + N_s^{\pi} + N_a^{\pi} = N$, where $N$ stands for a total number
of delocalized electrons in the fullerene
(four valence 2s$^2$2p$^2$ electrons from each carbon atom result in $N = 240$ in case of C$_{60}$).
The frequencies of the collective excitations are defined as
\cite{Lambin_Lukas_1992_PhysRevB.46.1794, Oestling_1993_EurophysLett.21.539, Lo_2007_JPhysB.40.3973}:
\begin{eqnarray}
\om_{s/a}^{\sigma+\pi} = \om_0 +
\left[ \frac{N_{s/a}^{\sigma+\pi}}{2(R_2^3 - R_1^3)} \left( 3 \mp \sqrt{1 + 8\xi^3} \right) \right]^{1/2} \ ,
\nonumber \\
\om_{s/a}^{\pi} =
\left[ \frac{N_{s/a}^{\pi}}{2(R_2^3 - R_1^3)} \left( 3 \mp \sqrt{1 + 8\xi^3} \right) \right]^{1/2} \ ,
\end{eqnarray}

\noindent where the signs '$-$' and '$+$' correspond to the symmetric and antisymmetric modes,
respectively,
and $\xi = R_1/R_2$ is the ratio of the inner to the outer radii.
The term $\om_0$ defines the free-electron picture threshold for the $(\sigma+\pi)$-plasmon
\cite{Oestling_1993_EurophysLett.21.539}.
Below $\om_0$, some of the valence electrons are treated as bound ones and, therefore, are not involved
in the formation of plasmon excitations. Following Ref. \cite{Oestling_1993_EurophysLett.21.539},
we use the threshold value $\om_0 = 14$ eV in the calculations.

\section{Computational Details}

In order to calculate the photoabsorption
spectrum of C$_{60}$ within the {\it ab initio} framework we
utilized a combination of various computer packages.
The \verb"Gaussian09" package \cite{g09} was used to optimize the geometry of the molecule.
The optimization procedure was performed by means of DFT within the local density approximation (LDA)
using the split-valence triple-zeta 6-311+G(d) basis set with an additional set of polarization and
diffuse functions.
To account for the exchange and correlation corrections, the Slater exchange functional
\cite{Kohn_Sham_1965_PhyRev.140.A1133} and the local Perdew functional
\cite{Perdew_Zunger_1981_PhyRevB.23.5048} were used.
The photoabsorption spectrum of the optimized system was obtained using the \verb"TDDFPT" module
\cite{Malcioglu_2011_CompPhysCommun.182.1744} of the \verb"QuantumEspresso" package
\cite{Giannozzi_2009_JPhysCondMat.21.395502}.
The optimized structure of the C$_{60}$ molecule was introduced into a supercell of
$20 \times 20 \times 20~{\rm \AA}$. 
Then, the system of Kohn-Sham equations was solved self-consistently for 240 valence electrons
of the fullerene to calculate the ground-state eigenvalues using a plane-wave approach
\cite{Giannozzi_2009_JPhysCondMat.21.395502}.
It should be noted that a similar approach was utilized in
\cite{Chen_Msezane_2012_EPJD.66.184, Chen_Msezane_2012_PhysRevA.86.063405} to study photoionization
of noble gas atoms encapsulated in C$_{60}$.
In the present calculations, we used a ultrasoft pseudopotential of the Rappe-Rabe-Kaxiras-Joannopoulos-type \cite{RRKJ_pseudopotential}, which substitutes real atomic orbitals in the core region with smooth
nodeless pseudo-orbitals \cite{Walker_2007_JChemPhys.127.164106}.
For the plane-wave calculations we used the kinetic energy cutoff of 30 Ry for the wave functions and
180 Ry for the electron densities.
The results obtained were validated by performing a series of calculations with different values of the
supercell size and the energy cutoff.

One should also mention that a pseudopotential local-density method, based on {\it ab initio} principles,
was also used earlier to calculate the electronic structure of solid C$_{60}$ \cite{Martins_1991_ChemPhysLett.180.457}.
The analysis performed showed that the occupied states of C$_{60}$ and some of the empty states can be
described as either $\sigma$ or $\pi$ states, which are formed due to the sp$^2$-hybridization of carbon
atomic orbitals.

\section{Results}

\subsection{Contribution of plasmon excitations}

\begin{figure}
\centering
\includegraphics[scale=0.58,clip]{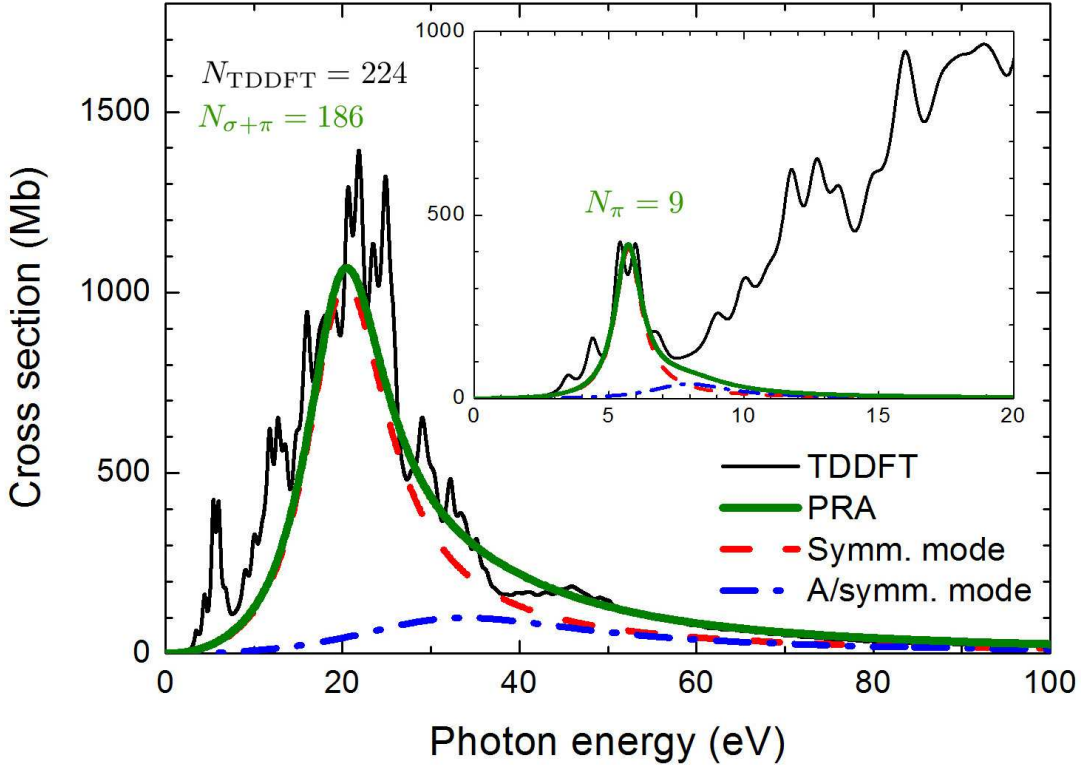}
\caption {
(Color online) The photoabsorption
cross section of C$_{60}$ calculated within the TDDFT method (thin black line)
and the plasmon resonance approximation (thick green line).
The curves, obtained within the classical approach, describe the dominating plasmon resonance, which is
formed due to collective oscillations of ($\sigma+\pi$) delocalized electrons of the system, and
a narrow low-energy peak below 10 eV (shown in the inset) which is attributed to the collective excitation of
only $\pi$-electrons.
Contributions of the symmetric and antisymmetric modes of the plasmons are shown by the
dashed (red) and dash-dotted (blue) lines, respectively.}
\label{figure1}
\end{figure}

Figure \ref{figure1} shows the photoabsorption
spectrum of C$_{60}$ calculated within the {\it ab initio}
and classical approaches in the photon energy region up to 100 eV.
The thin solid (black) line represents the results of TDDFT calculations within the LDA approach,
and the thick solid (green) one represents the contribution from the plasmon excitations.
The main resonant structure presented in Fig. \ref{figure1} is formed due to collective oscillations
of both $\sigma$- and $\pi$-electrons of the system, while a prominent peak in the low-energy region of
the spectrum (shown in the inset) is attributed to the collective excitation of only $\pi$-electrons.
The dashed (red) and dash-dotted (blue) lines show, respectively, contributions from the symmetric and
antisymmetric modes of the plasmons to the cross section.
The resonance frequencies, $\om_s$ and $\om_a$, for the two modes of the ($\sigma+\pi$)- and $\pi$-plasmons
as well as the corresponding widths, $\Gamma_s$ and $\Gamma_a$, are summarized in Table~\ref{table1}.
The width $\Gamma_s^{\sigma+\pi} = 11.4$ eV of the symmetric $(\sigma+\pi)$-plasmon mode
corresponds to the experimental values obtained from the photoionization and energy loss experiments
on neutral C$_{60}$ \cite{Hertel_1992_PhysRevLett.68.784, Mikoushkin_1998_PhysRevLett.81.2707}.
For the antisymmetric mode, we used the value $\Gamma_a^{\sigma+\pi} = 33.2$ eV which corresponds to
the widths of the second plasmon resonance obtained in the study of photoionization of
C$_{60}^{q+}$ ($q = 1-3$) ions \cite{Scully_2005_PhysRevLett.94.065503}.

\begin{table}
\centering
\caption{
Peak positions and the widths of the two modes of the ($\sigma+\pi$)- and $\pi$-plasmons used in the present
calculations. All values are given in eV.}
\begin{tabular}{p{3.5cm}p{1.0cm}p{1.0cm}p{1.0cm}p{1.0cm}}
\hline
                        & $\om_s$ & $\Gamma_s$ & $\om_a$ & $\Gamma_a$ \\
\hline
($\sigma+\pi$)-plasmon  &  20.3   &    11.4    &  33.5   &    33.2    \\
$\pi$-plasmon         &   5.8   &     1.2    &   7.9   &     3.5    \\
\hline
\end{tabular}
\label{table1}
\end{table}

The plasmon resonance approximation describes quite well the main features of the spectrum,
such as height, width and position of the plasmon resonance peaks.
The spectrum calculated within the TDDFT approach reveals a more detailed structure which is formed atop
the plasmon resonances and represents a series of individual peaks.
The oscillator strengths, calculated by means of TDDFT and within the plasmon resonance approximation in
the photon energy range up to 100 eV, are equal to 224 and 195, respectively.
Analysis of the plasmon contribution to the cross section shows that about 9 $\pi$-electrons are involved
in the low-energy collective excitation below 10 eV. This value corresponds to the experimentally
evaluated sum rule of the oscillator strength up to the ionization threshold of C$_{60}$,
$I_p \approx 7.6$ eV, which gives the value of 7.8 \cite{Berkowitz_1999_JChemPhys.111.1446}.

\begin{figure}
\centering
\includegraphics[scale=0.55,clip]{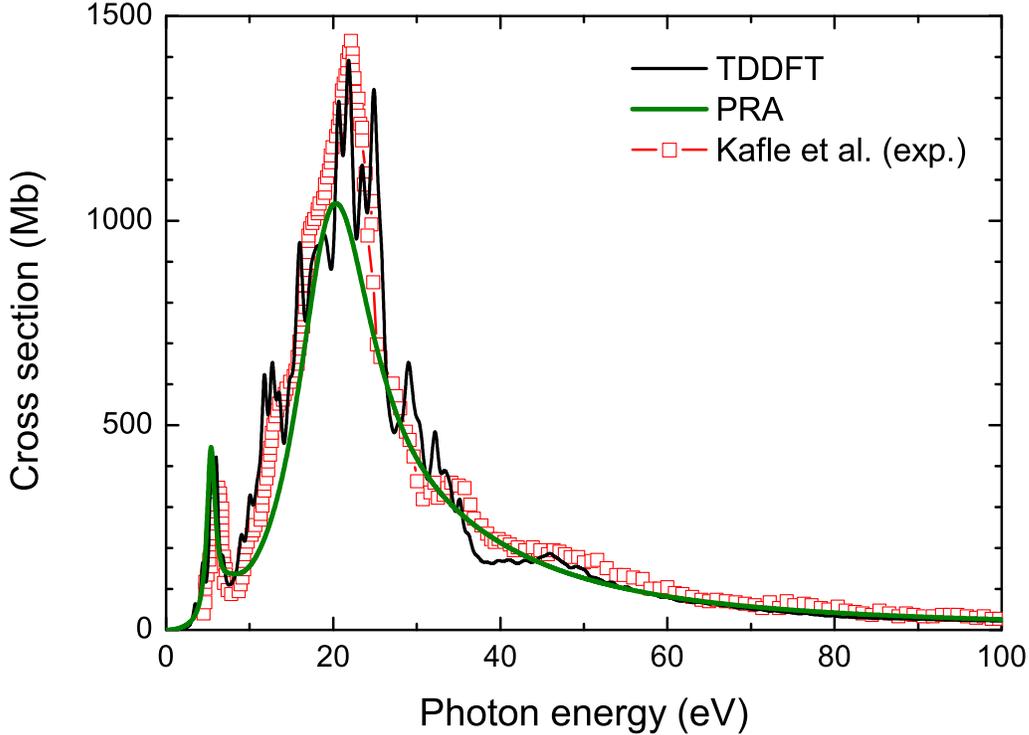}
\caption {
(Color online) The photoabsorption
cross section of C$_{60}$ calculated within the TDDFT method (thin black line)
and the plasmon resonance approximation (thick green line).
The curve, obtained within the classical approach, describes both the ($\sigma+\pi$)- and $\pi$-plasmons.
Theoretical curves are compared to the experimental data of Kafle {\it et al.}
\cite{Kafle_2008_JPhysSocJpn.77.014302}.}
\label{figure2}
\end{figure}

In Fig.~\ref{figure2}, the theoretical curves are compared to the results of recent experimental
measurements of photoabsorption of C$_{60}$ \cite{Kafle_2008_JPhysSocJpn.77.014302} (open squares).
The oscillator strength, calculated by means of TDDFT, is very close to the experimentally measured
value of 230.5 \cite{Kafle_2008_JPhysSocJpn.77.014302}.
It should be noted that the detailed structure of the spectrum, which is described within the TDDFT
approach, is not seen in the experimental curve due to a high operational temperature
of 500-700$^{\circ}$C \cite{Berkowitz_1999_JChemPhys.111.1446}.
In the experiments, the linewidths of single-electron excitations are broadened in the vicinity of the
main plasmon resonance due to the coupling of electron excitations with the vibrational modes of the
ionic background \cite{Gerchikov_2000_JPhysB.33.4905}.
The analysis, performed in the present study, shows that the plasmon resonance approximation gives an
adequate description of the experimental results.
A better agreement can be obtained if one incorporates the broadening of the linewidths
of single-electron excitations into the model and uses the calculated values of the widths.
This problem requires a separate detailed analysis and will be a subject for a further investigation.

\begin{figure}
\centering
\includegraphics[scale=0.5,clip]{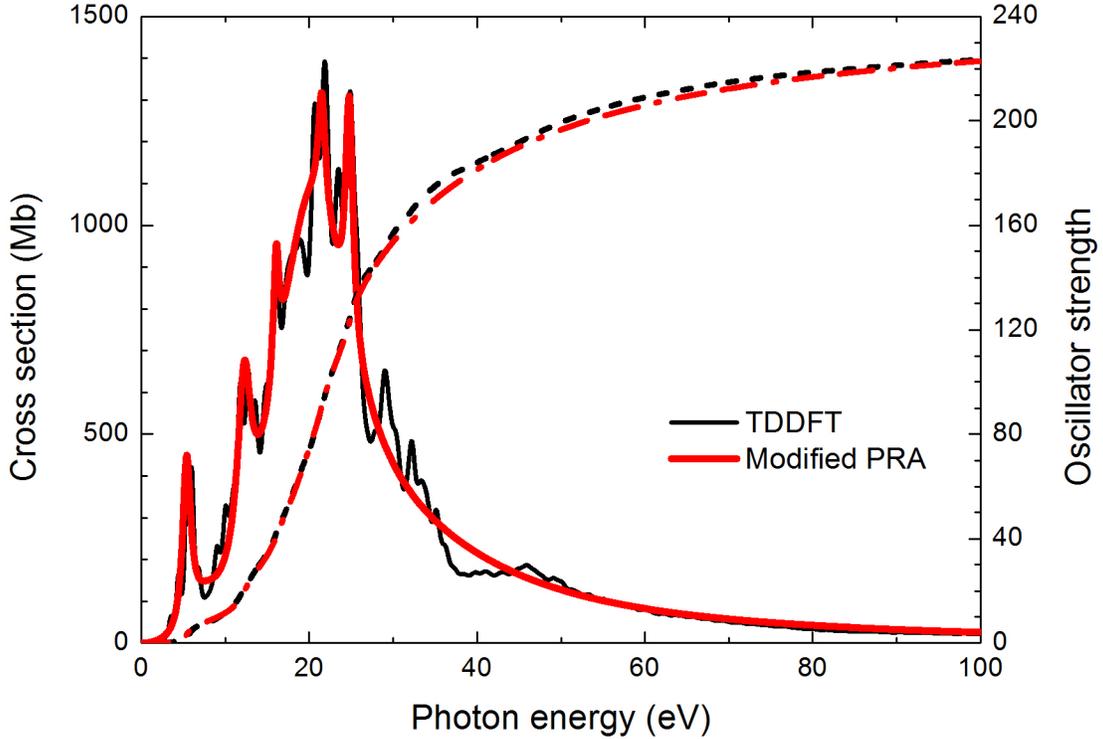}
\caption {
(Color online) The photoabsorption
cross section of C$_{60}$ calculated within the TDDFT approach (thin black line)
and the modified plasmon resonance approximation (thick red line), which estimates in a simple way
the single-particle contribution (see Eq.~(\ref{CS_modified})).
Dashed (black) and dash-dotted (red) lines represent the integrated oscillator strengths calculated
within the TDDFT-based approach and the modified plasmon resonance approximation, respectively.}
\label{figure3}
\end{figure}

The difference between the oscillator strengths, calculated within the TDDFT approach and the plasmon
resonance approximation, is due to the contribution from single-particle excitations, which are neglected
in the model.
This contribution can be estimated and added to the model-based results by describing individual peaks
in the TDDFT-based spectrum by a number of Lorentzian functions. Then, the total cross section reads as:
\begin{equation}
\sigma(\om) = \sigma_{\rm pl}(\om) + \frac{4\pi \om^2}{c}
\sum_j
{ f_j\,\Gamma_j \over \bigl(\om^2-\om_j^2\bigr)^2+ \om^2\Gamma_j^2} \ ,
\label{CS_modified}
\end{equation}

\noindent where $\sigma_{\rm pl}(\om)$ is the contribution from the ($\sigma+\pi$)- and $\pi$-plasmons,
defined by eq.~(\ref{CS_plasmon}), the index $j$ represents the number of Lorentzian functions used,
and $f_j$ stands for the summarized oscillator strength of one or several individual peaks in the TDDFT
spectrum, which are modeled by a single Lorentzian function.
In order to estimate the single-particle contribution to the cross section in the range from 10 to 25 eV,
we introduced 4 Lorentzian profiles, as it is shown in Fig.~\ref{figure3}.
The summarized curve, which comprises the contribution from the plasmon excitations and a simple estimate
for the single-particle excitations, is shown by the thick red line.
Figure~\ref{figure3} represents also the comparison of the integrated oscillator strengths, calculated
within the TDDFT-based approach (dashed black curve) and the modified plasmon resonance approximation defined
by Eq.~(\ref{CS_modified}) (dash-dotted red curve).
The analysis of the two curves demonstrates that such a simple estimate for the single-particle contribution
covers the difference in the oscillator strengths calculated within the TDDFT and the model approaches.
The oscillator strengths, calculated by means of TDDFT and within the modified plasmon resonance approximation
in the photon energy range up to 100 eV, are equal to 224 and 223.5, respectively.

\subsection{Quantum nature of individual peaks}

\begin{figure}
\centering
\includegraphics[scale=0.49,clip]{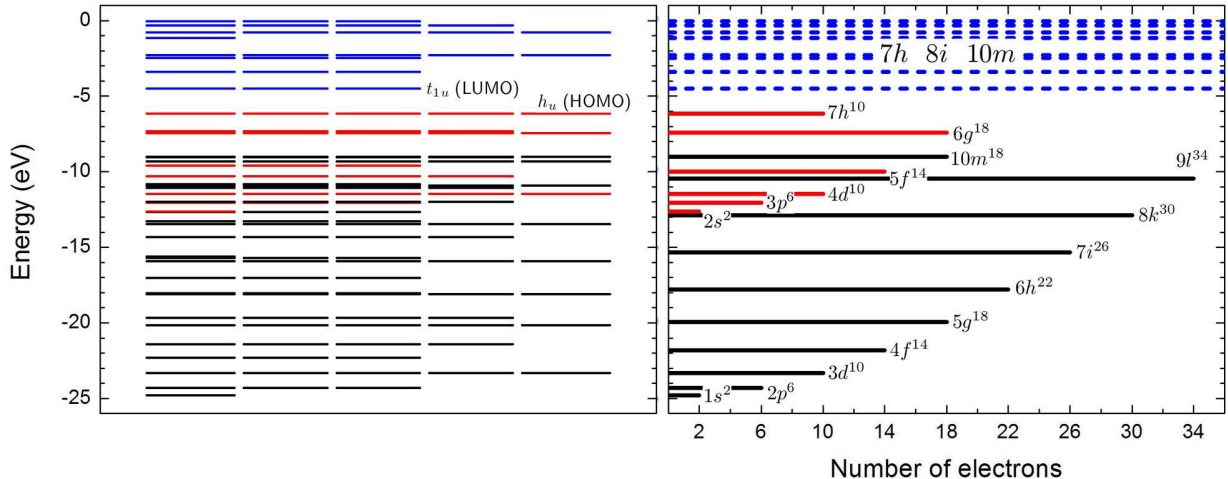}
\caption {
(Color online)
Left panel: The ground-state electronic structure of C$_{60}$ obtained within the {\it ab initio}
framework accounting for the real $I_h$ symmetry of the molecule.
Each line corresponds to one molecular orbital which accommodates (or may accommodate) two electrons.
Black and red lines (in the range $-6 \dots -25$ eV) represent the MOs, which are occupied by 240
delocalized valence electrons of C$_{60}$. Blue lines (above $-5$ eV) represent virtual bound states.
Right panel: Electronic structure of the corresponding spherically symmetric $nl$-orbitals.
The horizontal lines indicate the occupation numbers for each orbital and correspond to the summarized
number of electrons which occupy the corresponding MOs.}
\label{figure4}
\end{figure}

Let us focus now on the origin of individual peaks which are formed in the photoionization spectrum of C$_{60}$
atop the plasmon resonances.
These peaks can be assigned to discrete transitions between particular molecular orbitals (MOs).
The C$_{60}$ fullerene belongs to the icosahedral ($I_h$) symmetry group, therefore its MOs can be
classified according to the $I_h$ irreducible representations.
The icosahedral symmetry allows the maximum orbital degeneracy equal to five.
Thus, the MOs can be singly ($a_g$, $a_u$),
triply ($t_{1g}$, $t_{1u}$), ($t_{2g}$, $t_{2u}$), fourfold ($g_g$, $g_u$) and fivefold ($h_g$, $h_u$)
degenerated. The subscripts '$g$' and '$u$' denote, respectively, symmetric (''gerade'') and antisymmetric
(''ungerade'') MOs with respect to the center of inversion of the molecule.
Due to the quasispherical structure of the molecule, the MOs can be expanded in terms of spherical
harmonics in the angular momentum $l$ \cite{Saito_1991_PhysRevLett.66.2637} (see Fig.~\ref{figure4}).
Thus, the innermost $a_g$, $t_{1u}$ and $h_g$ MOs in the $I_h$ symmetry represent, respectively,
the $s$, $p$ and $d$ orbitals, which correspond to $l = 0, 1$ and $2$.
The orbitals which correspond to higher angular momenta are constructed as a combination of several MOs.
The correspondence between the MOs of C$_{60}$ and the spherically
symmetric orbitals with a given value of angular momentum $l$ is given in Table~\ref{table2}.

\begin{table}
\centering
\caption{
Molecular orbitals occupied by delocalized electrons of C$_{60}$ (left column) and the corresponding
spherically symmetric orbitals which are obtained by the expansion of real MOs in terms of spherical
harmonics in the angular momentum $l$ (right column).}
\begin{tabular}{p{5cm}p{0.2cm}p{1.5cm}}
\hline
$a_g$                                &  $s$  &  ($l = 0$)  \\
$t_{1u}$                             &  $p$  &  ($l = 1$)  \\
$h_g$                                &  $d$  &  ($l = 2$)  \\
$g_u + t_{2u}$                       &  $f$  &  ($l = 3$)  \\
$h_g + g_g$                          &  $g$  &  ($l = 4$)  \\
$h_u + t_{1u} + t_{2u}$              &  $h$  &  ($l = 5$)  \\
$a_g + t_{1g} + g_g + h_g$           &  $i$  &  ($l = 6$)  \\
$h_u + t_{1u} + t_{2u} + g_u$        &  $k$  &  ($l = 7$)  \\
$h_g + g_g + t_{2g} + h_g$           &  $l$  &  ($l = 8$)  \\
$g_u + h_u + g_u + t_{1u} + t_{2u}$  &  $m$  &  ($l = 9$)  \\
\hline
\end{tabular}
\label{table2}
\end{table}

In the spherical representation of C$_{60}$, the delocalized electrons are considered as moving in
a spherically symmetric central field.
Therefore, one can construct the ground-state electronic configuration described by the unique set of
quantum numbers $\{n, l\}$ where $n$ and $l$ are the principal and orbital quantum numbers, respectively
\cite{Yabana_Bertsch_1993_PhysScr.48.633}:
%
\begin{center}
$1s^2 2p^6 3d^{10} 4f^{14} 5g^{18} 6h^{22} 7i^{26} 8k^{30} 9l^{34} 10m^{18}$ \\
$2s^2 3p^6 4d^{10} 5f^{14} 6g^{18} 7h^{10}$.
\end{center}

\noindent The superscripts indicate the occupation numbers for each spherically symmetric orbital and
correspond to the summarized number of electrons which occupy the corresponding MOs
(see the left and right panels of figure~\ref{figure4}).
One may consider the icosahedral symmetry of C$_{60}$ as a perturbation of the spherical one, so
the correspondence between the real MOs and the spherically symmetric $nl$-orbitals can be explained in
terms of splitting of the latter ones due to reduction of the symmetry.

\begin{figure}
\centering
\includegraphics[scale=0.43,clip]{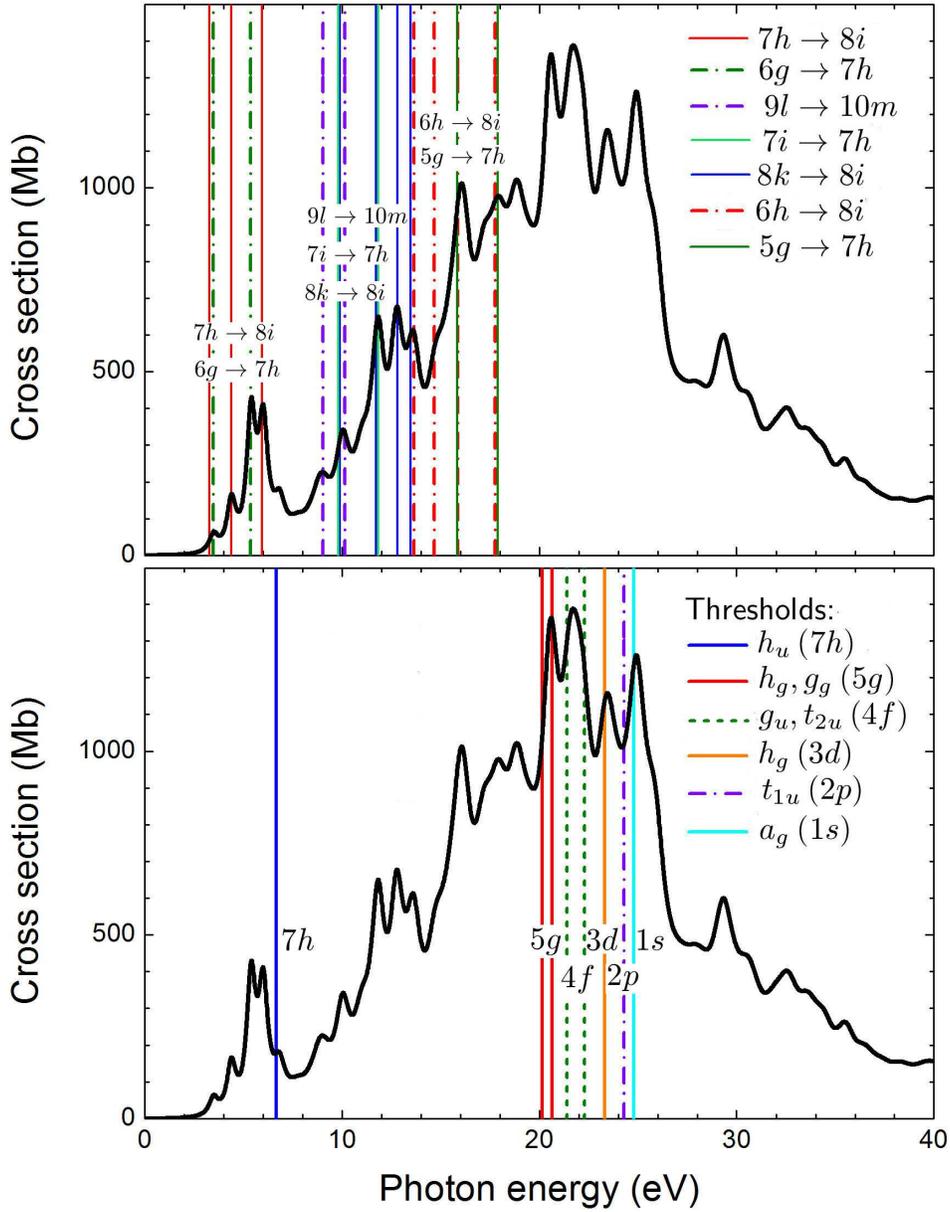}
\caption {
(Color online)
Upper panel: Excitation energies of the optically allowed discrete transitions to the virtual bound
states which correspond in the spherical representation of C$_{60}$ to the unoccupied or partially
occupied $7h$, $8i$ and $10m$ orbitals (see the text for details).
Lower panel: Ionization thresholds of the highest occupied molecular orbital, $h_u$
(partially occupied $7h$ orbital), as well as of the following innermost valence orbitals:
$h_g$ and $g_g$ ($5g$), $g_u$ and $t_{2u}$ ($4f$), $h_g$ ($3d$), $t_{1u}$ ($2p$), and $a_g$ ($1s$).
In the LDA calculations, the ionization thresholds for the HOMO ($h_u$), and the innermost valence ($a_g$)
molecular orbitals are 6.65 eV and 24.8 eV, respectively.}
\label{figure5}
\end{figure}

Using the \verb"Gaussian09" package \cite{g09}, we calculated a number of virtual bound states
of C$_{60}$ ($t_{1u}, t_{1g}, t_{2u}, h_g, a_g, h_u, g_u$) which can be assigned in the spherical
representation to the unoccupied or partially occupied $7h$, $8i$ and $10m$ orbitals
(see the blue dashed lines in Fig.~\ref{figure4}).
The optically allowed discrete transitions should result in the change of the MO's symmetry ($g \leftrightarrow u$)
or satisfy the $l \to l \pm 1$ selection rule within the spherical representation.
We listed all possible optically allowed discrete transitions and calculated the corresponding
transition energies.
The results of this analysis are summarized in the upper panel of Fig.~\ref{figure5}.
The peaks in the TDDFT spectrum can be assigned to
$5g, 6g \to 7h$; $7i \to 7h$; $6h, 7h \to 8i$; $8k \to 8i$ and $9l \to 10m$ transitions
(in order to simplify the analysis we use here the spherical representation of orbitals).
The discrete transitions are shown in the upper panel of Fig.~\ref{figure5} by thin solid
and dash-dotted vertical lines.
The six lowest optically allowed $\pi - \pi^*$ excitations
($h_u \to t_{1g}$, $h_g \to t_{1u}$, $h_u \to h_g$, $g_g \to t_{2u}$, $h_g \to t_{2u}$, and $h_u \to g_g$)
 \cite{Saito_1991_PhysRevLett.66.2637} from 3 to 6 eV
correspond to $7h \to 8i$ and $6g \to 7h$ transitions (see solid red and dash-dotted green lines),
which are involved in the formation of the $\pi$-plasmon.
The $9l \to 10m$, $7i \to 7h$ and $8k \to 8i$ transitions
(violet, aquamarine and blue lines, respectively) result in the formation of individual peaks
in the region from 9 to 14 eV, which are formed atop the ($\sigma+\pi$)-plasmon excitation.
Features in the energy range from 14 to 18 eV are assigned to the single-electron $6h \to 8i$ and
$5g \to 7h$ transitions from the lower-lying $5g$ and $6h$ orbitals.
Thus, accounting for the optically allowed discrete transitions it is possible to reveal a detailed
structure of the photoionization spectrum of C$_{60}$ up to 18 eV.
However, one can extend the analysis and characterize a number of subsequent peaks.

In the lower panel of Fig.~\ref{figure5}, vertical lines represent the ionization thresholds of
several particular orbitals.
The highest-occupied molecular orbital ($h_u$) of C$_{60}$ corresponds in the spherical
representation to the partially filled $7h$ orbital.
In the present calculations performed within the LDA approach, the $h_u$ ionization threshold is 6.65 eV
(solid blue line) which is slightly lower than the experimentally measured ionization potential of C$_{60}$,
approximately equal to 7.6 eV \cite{Hertel_1992_PhysRevLett.68.784}.
Since there are no discrete optical transitions with the energy above 20 eV, a series of peaks, arising
between 20 and 25 eV, can be assigned to the ionization of the innermost fullerene MOs
(the corresponding $nl$-orbitals are given in parentheses),
namely $h_g$ and $g_g$ ($5g$), $g_u$ and $t_{2u}$ ($4f$), $h_g$ ($3d$), $t_{1u}$ ($2p$), and $a_g$ ($1s$).
The calculated ionization thresholds of these orbitals are shown in the lower panel of Fig.~\ref{figure5}.
Within the LDA approach, the threshold of the innermost molecular orbital ($a_g$) equals to 24.8 eV
(solid cyan line).

The information, which can be obtained within the {\it ab initio} framework, allows one to reveal clearly the
origin of the individual peaks in the photoionization spectrum of C$_{60}$ for the photon energies up to 25 eV.
The nature of several subsequent peaks, located at about and above 30 eV, cannot be explored by the
{\it ab initio} approach and should be investigated by means of the model one.
One may suppose that these peaks should be caused by excitation of particular molecular orbitals to the
continuum.


\section{Conclusion}

To conclude, we have performed a detailed theoretical analysis of the photoabsorption spectrum of the C$_{60}$
fullerene and revealed the contributions coming out from single-particle and collective electron excitations.
On the basis of the {\it ab initio} calculations, performed within the time-dependent density-functional
theory framework, we elucidated the origin of various quantum phenomena which manifest themselves 
atop the plasmon excitations.

We have demonstrated that the individual peaks lying below and above the ionization threshold,
in the photon energy region from 3 to 18 eV, are due to optically allowed discrete
transitions which are formed atop the collective excitation of $\sigma+\pi$ delocalized electrons.
In the spherical representation of molecular orbitals in terms of angular momentum $l$, these transitions
correspond to the following ones: $5g, 6g \to 7h$; $7i \to 7h$; $6h, 7h \to 8i$; $8k \to 8i$, and $9l \to 10m$.
The analysis performed shows that the peaks in the vicinity of the dominating ($\sigma+\pi$)-plasmon resonance,
between 20 and 25 eV, are caused by the ionization of the innermost fullerene orbitals, namely
$h_g$ and $g_g$ ($5g$), $g_u$ and $t_{2u}$ ($4f$), $h_g$ ($3d$), $t_{1u}$ ($2p$), and $a_g$ ($1s$).
Finally, the peaks around 30 eV and above cannot be explained within the pure {\it ab initio} framework due to
lack of information, therefore one should use some model approaches to explore the origin of these peaks.


The broad resonance peak in the photoabsorption
cross section of C$_{60}$ is formed due to the ($\sigma+\pi$)-plasmon.
A number of discrete excitations, lying below and just above the ionization threshold of C$_{60}$, are attributed to a
collective excitation of delocalized $\pi$-electrons.
The spectrum calculated within the model approach is in good agreement with that obtained by means of the more
sophisticated TDDFT method and corresponds to the results of experimental measurements.
Therefore, the plasmon resonance approximation, utilized in the present work, represents a useful tool for the
interpretation of the {\it ab initio} calculations and experimental measurements.
A better agreement of the model approach with the experimental data can be obtained if one incorporates the broadening
of the linewidths of single-electron excitations into the model and uses the calculated values of the widths.
This problem will be a subject for a further investigation.

\section*{Acknowledgements}

The authors acknowledge the Frankfurt Center for Scientific Computing for the opportunity of carrying out
complex resource-demanding calculations.


\end{document}